\documentclass[aps,prl,showpacs,amsmath,amssymb,amsfonts,twocolumn,
superscriptaddress,floatfix,footinbib]{revtex4}

\usepackage{subfigure}
\usepackage{graphicx}
\usepackage[flushmargin]{footmisc}

\newcommand{\x}{\vec{x}}

\newcommand{\kv}{\vec{k}}

\newcommand{\wL}{\omega_{TM}}
\newcommand{\wT}{\omega_{TE}}

\newcommand{\M}{\mathbf{\hat{M}}}

\newcommand{\rv}{\vec{r}}

\newcommand{\T}{\mathbf{\hat{T}}}

\begin{document}

\title{Excitation of Vortices in Semiconductor Microcavities}

 \author{T C H Liew}
 \author{A V Kavokin}
 \author{I A Shelykh}
 \affiliation{School of Physics and Astronomy, University of
 Southampton, Highfield, Southampton SO17 1BJ, UK}

\pacs{71.36.+c, 42.65.-k, 03.75.Kk}
\date{\today}

\begin{abstract}

\noindent We predict that the transverse electric-magnetic
polarization splitting of exciton-polaritons allows a simple
technique for the generation of polariton vortices of winding number
2 in semiconductor microcavities. The vortices can be excited by
circularly polarized light having no orbital angular momentum and
observed as phase vortices in the opposite circular polarization or
directly as linear polarization vortices. The prediction is
explained by a simplified analytical linear model and shown fully
with a numerical model based on the Gross-Pitaevskii equations,
which includes the non-linear effects of polariton-polariton
interactions.
\end{abstract}

\maketitle


{\bf Introduction.} The creation and evolution of vortices in atomic
Bose-Einstein condensates (BECs) has attracted much
attention~\cite{Anglin,Williams} since the experimental realization
of a dilute atomic BEC~\cite{Anderson} just over a decade ago. It is
no coincidence that vortices are at the heart of our understanding
of superfluidity, forming spontaneously in type II
superconductors~\cite{Abrikosov} and
superfluids~\cite{Onsager,Feynman}. In atomic BECs vortices can be
optically created using light modes with orbital angular momentum
(Laguerre-Gaussian modes) and exploiting the recent technology of
ultraslow light pulses to couple light and matter
fields~\cite{Dutton}. Vortices written by light fields are stored in
the atomic condensate and can later be re-written onto light fields.

\begin{figure}[h]
        \centering
        \includegraphics[width=8.116cm]{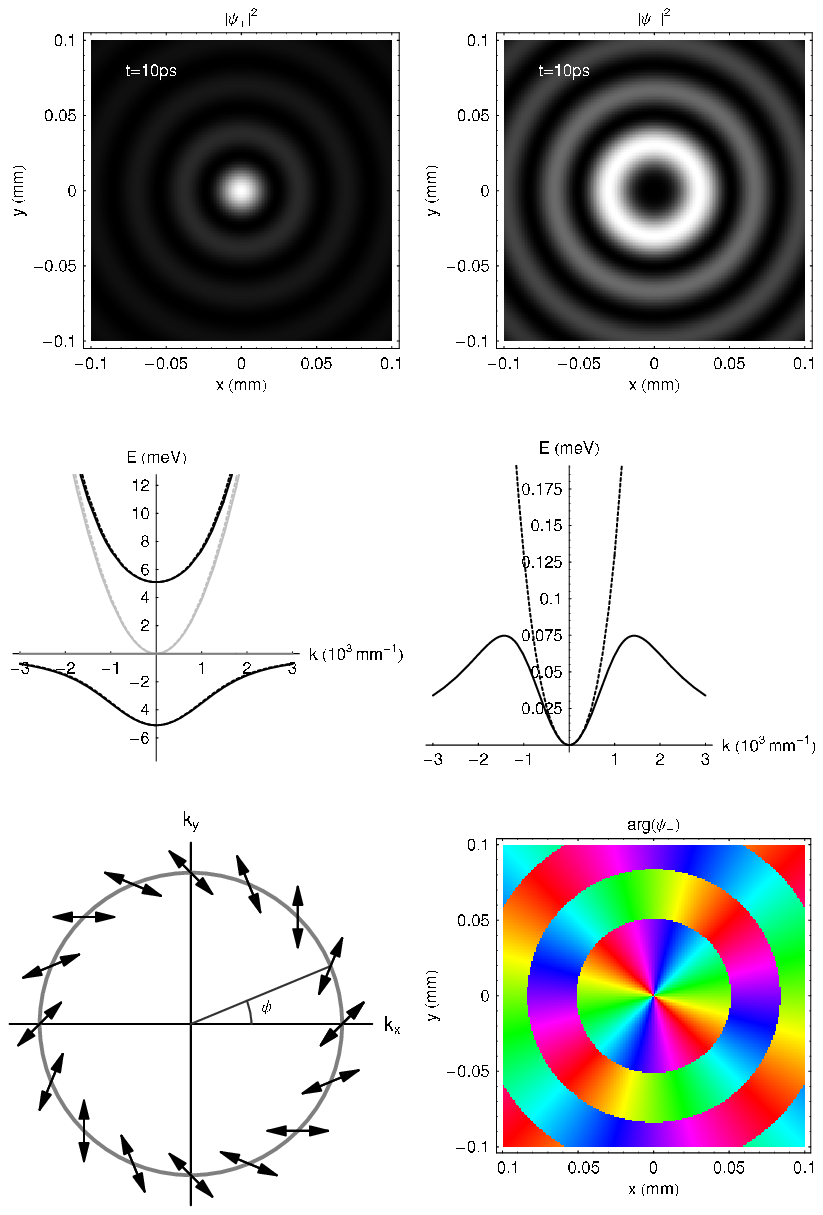}
    \caption{({\it colour}) (Top:) Plots of the absolute values of $\Psi_{+}$
    (left) and $\Psi_{-}$ (right) for $k_p=100$mm$^{-1}$. Note that the gray-scales are not the same; the cross-circular component is much weaker (depending strongly on the value of the TE-TM splitting at
    $k_p$). (Center Left:) The dispersion of photons and excitons (gray) and polaritons (black) for an exciton-photon coupling energy (vacuum field Rabi splitting) of $5.1$meV. The TE-TM splitting of
    the photon is given by the different effective masses
    of TE and TM branches. We chose $m_t=10^{-5}m_e$
    and $m_l=0.95m_t$ where $m_e$ is the free electron mass. (Center Right:) The parabolic TE-TM splitting of the
    photon causes an TE-TM splitting of the lower (solid) and upper
    branch (dashed) polaritons. (Bottom Left:) The linear polarization vortex. Arrows indicate the linear polarization along the circle $k=k_p$ after a
    time $t=\frac{\pi}{2(\wL-\wT)}$.
    (Bottom Right:) The phase of $\Psi_{-}$ changes by $4\pi$ around the origin.}
    \label{fig:AnalyticVortex}
    \end{figure}

In semiconductor systems, the high temperature BEC of the half-light
half-matter quasiparticles called exciton-polaritons has been
reported recently~\cite{Kasprzak}. The condensation takes place in a
system of several quantum wells sandwiched between two distributed
Bragg reflectors, commonly known as a microcavity. Many unique
properties of polaritons~\cite{CavityPolaritons} arise from their
non-parabolic in-plane dispersion (center left of Fig.
\ref{fig:AnalyticVortex}). No vortices of polaritons have been
observed so far, although superfluidity in polariton systems has
been largely discussed theoretically~\cite{Carusotto,Shelykh2006}.
Here we show that a unique fine structure of polaritons in
microcavities allows for a very simple mechanism of exciting
vortices using conventional circularly polarized light beams having
zero orbital angular momentum.

It is well known that transverse electric (TE) and magnetic (TM)
normal modes of a cavity are non-degenerate for finite in-plane
wavevectors. Polaritons also possess a polarization dependent TE-TM
energy splitting~\cite{LTSplitting}, which was previously proposed
to allow the optical spin Hall effect~\cite{OpticalSHE}. We will
show that this splitting can cause an initially circularly polarized
polariton distribution to form a phase vortex in the cross-circular
polarization, characterized by a phase singularity around which the
phase changes by $4\pi$. A vortex structure also appears in the
linear polarization pattern.

This linear effect is a general property of all planar cavities, not
only those with quantum wells, and could be useful for creating
light modes with angular momentum. Such light modes can have
interesting applications to quantum information~\cite{Mair}. Our
next step is to take into account the many-body polariton-polariton
interactions, which play a crucial role in a number of microcavity
processes where large polariton populations are
involved~\cite{PolaritonLaser,ParametricAmplifier}.

Polariton-polariton interactions can be accounted for using the
zero-range interaction and mean-field approximations that lead to
the Gross-Pitaevskii equations. In Ref.~\cite{Carusotto} the
polarization independent spatial dynamics of interacting polariton
systems was derived by numerically solving the single component
(scalar) Gross-Pitaevskii equations. For us, the inclusion of
polarization is not only important due to effects caused by TE-TM
splitting; polariton-polariton interactions themselves were
previously shown to be spin-anisotropic, which led to effects such
as self-induced Larmor precession of elliptical polarizations and
inversion of linear polarizations~\cite{LarmorPrecession}.

Taking into account polarization, polaritons represent a two
component Bose gas. Vortices in two-component bosonic systems are
closely related to Skyrmions~\cite{Ruostekoski}, which makes
microcavities a potentially suitable solid state system for Skyrmion
observation.


{\bf Linear Model.} We will first outline a simple analytic model to
show how vortices can be generated in microcavities. First we
consider a lower branch polariton state characterized by the
in-plane wavevector $\kv$ rotated by an angle $\phi$ with respect to
the $x-$axis. A circularly polarized state is a linear combination
of TM and TE eigenstates, given by:

\begin{equation}
\left[\begin{array}{c}\psi_{TM0}\\\psi_{TE0}\end{array}\right] =\M
\left(\begin{array}{c}1\\0\end{array}\right)
\end{equation}

\begin{equation}
\M=\frac{1}{\sqrt{2}}\left(\begin{array}{cc} e^{i\phi}&
e^{-i\phi}\\i e^{i\phi}&-i e^{-i\phi}\end{array}\right)
\end{equation}

\noindent The evolution of the initial state in the circular basis
is:

\begin{align}
\left(\begin{array}{c}\psi_{+}\\\psi_{-}\end{array}\right)&=\M^{-1}
\left[\begin{array}{c}\psi_{TM0}e^{i\wL t-t/\tau}\\\psi_{TE0}e^{i\wT
t-t/\tau}\end{array}\right]\notag\\&=e^{i\frac{\wL+\wT}{2}t-t/\tau}
\left(\begin{array}{c}\cos{\left(\frac{\wL-\wT}{2}t\right)}\\i e^{2
i\phi}\sin{\left(\frac{\wL-\wT}{2}t\right)}\end{array}\right)\label{eq:VortexA}
\end{align}

\noindent where $\wL$ and $\wT$ are the frequencies of the two
eigenstates and $\tau$ accounts for the life-time of polaritons. The
k-dependent energy splitting between $\wL$ and $\wT$ is shown in the
center right plot of Fig. \ref{fig:AnalyticVortex}. From Eq.
(\ref{eq:VortexA}) we see that the phase of the cross-circularly
polarized component ($\psi_{-}$) is angle dependent. So far we have
only considered a single state with wavevector $\kv$. To obtain the
field distribution from multiple wavevector states we should take
the convolution integral with a pump distribution, $f(k,\phi,t)$:

\begin{align}
&\left(\begin{array}{c}\Psi_{+}(\rv,t)\\\Psi_{-}(\rv,t)\end{array}\right)=
\int_{-\infty}^t dt^\prime
e^{-(t-t^\prime)/\tau}\int_0^{2\pi}{d\phi}\int_0^{\infty}kdk
\notag\\&\hspace{5mm}\times e^{i\frac{\wL+\wT}{2}(t-t^\prime)}
e^{ikr\cos(\theta-\phi})f(k,\phi,t)\notag\\
&\hspace{5mm}\times\left(\begin{array}{c}\cos{\left(\frac{\wL-\wT}{2}(t-t^\prime)\right)}\\i
e^{2
i\phi}\sin{\left(\frac{\wL-\wT}{2}(t-t^\prime)\right)}\end{array}\right)
\end{align}

\noindent where $\theta$ is the angle between the radius vector
$\rv$ and the $x-$axis. For simplicity we consider excitation by a
delta function pulse $f(k,\phi,t)=A\delta(k-k_p)\delta(t)$. This
gives:

\begin{align}
\left(\begin{array}{c}\Psi_{+}(\rv,t)\\\Psi_{-}(\rv,t)\end{array}\right)=&2\pi
A k_p
e^{i\frac{\wL+\wT}{2}t-t/\tau}\notag\\
&\times\left(\begin{array}{c}J_0(k_pr)\cos{\left(\frac{\wL-\wT}{2}t\right)}\\-i
e^{2
i\theta}J_2(k_pr)\sin{\left(\frac{\wL-\wT}{2}t\right)}\end{array}\right)\label{eq:VortexB}
\end{align}

\noindent where $J_n$ denotes the $n$th order Bessel function of the
first kind. The amplitude and phase of $\Psi_{+}$ and $\Psi_{-}$ are
plotted in Fig. \ref{fig:AnalyticVortex}. It shows that the phase of
the cross-circular component changes by $4\pi$ around the origin. In
other words we have found a vortex of winding number 2. Note also
the role of oscillatory terms in Eqs. (\ref{eq:VortexA}) and
(\ref{eq:VortexB}): they describe the beats between the two spin
states of polaritons. Initially, all polaritons lie in the spin-up
state and there is no vortex. After a quarter of the period
$T=\frac{2\pi}{\wL-\wT}$ all polaritons lie in the spin-down state
and the vortex appears. Another quarter period later polaritons
return to the spin-up state and the vortex disappears. At the
intermediate time $t=\frac{T}{8}$ the polaritons are linearly
polarized, forming a polarization vortex as sketched in the bottom
left plot of Fig. \ref{fig:AnalyticVortex}.

{\bf Non-Linear Model.} In this section we use a numerical model to
include the effects of polariton-polariton interactions and the
finite extent of any real pulse in space and time. We will use the
Gross-Pitaevskii equations, which are a familiar tool for the
treatment of dilute Bose-gases~\cite{Gross1961,Pitaevskii1961}. For
a coherent ensemble of polaritons the set of coupled
Gross-Pitaevskii equations is written:
    \begin{align}
    i\hbar\frac{\partial\chi_i}{\partial t}&=\T^\chi_{ij}\chi_j+V\phi_i+(\alpha_1-\alpha_2)|\chi_i|^2\chi_i+\notag\\
    &\hspace{5mm}+\alpha_2\chi_j^*\chi_j\chi_i-\frac{i\hbar}{2\tau_\chi}\chi_i\hspace{10mm}
    \label{eq:GPchi}
    \end{align}
    \begin{align}
    i\hbar\frac{\partial\phi_i}{\partial t}&=\T^\phi_{ij}\phi_j+V\chi_i+f_i(\x,t)-\frac{i\hbar}{2\tau_\phi}\phi_i
    \label{eq:GPphi}
    \end{align}
\noindent (we sum over the index $j$). Equations (\ref{eq:GPchi},
\ref{eq:GPphi}) are obtained by the mean field treatment, valid at
low temperatures and densities where the thermal and quantum
depletion of the condensate has only a small effect. $\chi_i$ and
$\phi_i$ are wavefunctions representing the excitonic and photonic
components of the upper and lower branch polaritons. The indices
$i,j$ can take the values $+$ or $-$, representing the two circular
polarizations. The operators $\T^\chi$ and $\T^\phi$ are the kinetic
energy operators of uncoupled excitons and photons, which have
parabolic dispersion with effective masses $m_\chi$ and $m_\phi$:
    \begin{align}
    \T^{\chi}=&-\frac{\hbar^2}{2m_\chi}\left(
    \begin{array}{cc}
    \partial_{xx}+\partial_{yy}&0\\
    0&\partial_{xx}+\partial_{yy}
    \end{array}
    \right)
    \end{align}
    \begin{equation}
    \T^{\phi}=-\frac{\hbar^2}{2m_\phi}\left(
    \begin{array}{cc}
    \partial_{xx}+\partial_{yy}&\zeta\left(\partial_x-i\partial_y\right)^2\\
    \zeta\left(\partial_x+i\partial_y\right)^2&\partial_{xx}+\partial_{yy}
    \end{array}
    \right)
    \end{equation}
\noindent We have chosen an energy scale such that the energy of the
photonic component is zero at zero in-plane wavevector and consider
the case where there is no detuning between the exciton and photon
modes. The parameter $\zeta$ allows for a (parabolic) TE-TM
splitting of the photonic component in the microcavity (we neglect
the much smaller splitting of the excitonic component).

The coupling between excitons and photons is described by the Rabi
splitting of the cavity, $2V$, which is related to the quality
factor and the exciton decay
rate~\cite{CavityPolaritons,VCouplingConstant}. The non-linear terms
in Eq. (\ref{eq:GPchi}) describe Coulomb interactions between the
excitonic components where $\alpha_1$ describes the parallel spin
configuration and $\alpha_2$ describes the anti-parallel
configuration. In Eq. (\ref{eq:GPphi}) the term $f_i(\x,t)$
represents the optical pumping of the microcavity. For a coherent
Gaussian pump pulse exciting the lower polariton branch at the
wavevector $\vec{k}_0$ and energy $E_p$ at time $t_p$ the pump in
reciprocal space is:
    \begin{equation}
    f_i(\vec{k},t)=A_i e^{-iE_p
    t/\hbar}\frac{\Gamma e^{-(\vec{k}-\vec{k_0})^2/\Delta k^2} e^{-(t-t_p)^2/\Delta
    t^2}}{\left(E_{LP}(\vec{k})-E_p-i\Gamma)\right)}\label{eq:Pump}
    \end{equation}
\noindent $A_i$ represent the amplitudes of the two circular
polarizations of the pump pulse. $\Delta k$ and $\Delta t$ define
the width of the pump in reciprocal space and time. The fraction
$\Gamma/ \left(E_{LP}(\vec{k})-E_p-i\Gamma)\right)$ is the linear
susceptibility of a system of Lorentz oscillators where $\Gamma$ is
the homogeneous oscillator (HWHM) linewidth~\cite{HaugKoch}. In Eq.
(\ref{eq:GPphi}) the pumping term $\vec{f}(\x,t)$ is the inverse
Fourier transform of $\vec{f}(\vec{k},t)$. $E_{LP}(\vec{k})$ is the
bare dispersion of the lower polariton branch, which is calculated
by finding the eigenvalues of the linear Hamiltonian (without
interaction terms) in the exciton-photon
basis~\cite{PolaritonDispersion}. Finally, $\tau_\chi$ and
$\tau_\phi$ are the lifetimes of the excitonic and photonic
components, which account for the inelastic scattering and radiative
decay of polaritons respectively.

    \begin{figure}[h]
        \centering
        \includegraphics[width=8.116cm]{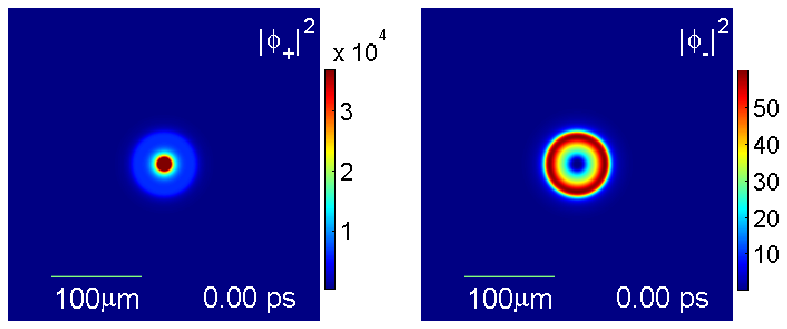}
        \\
        \includegraphics[width=8.116cm]{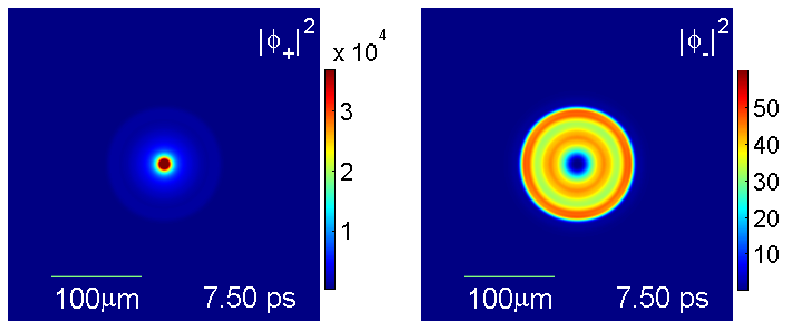}
        \\
        \includegraphics[width=8.116cm]{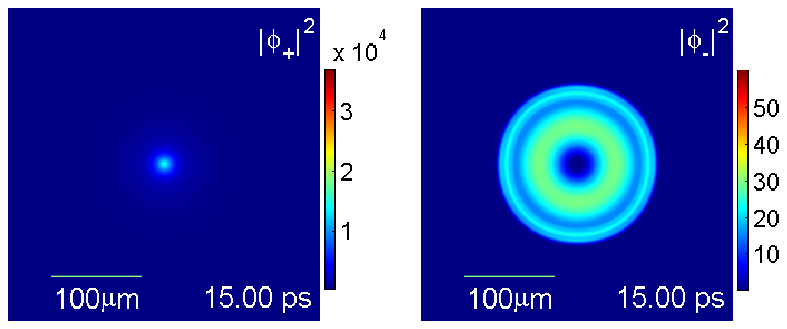}
    \caption{({\it colour}) Plots of the absolute values of the photonic
    wavefunction squared, $|\phi_i|^2$, at different times for the co-circular (left) and cross-circularly polarized
    components (right). The pulse was centered at $t_p=0$ps and had widths: $\Delta t=10$ps and $\Delta L=2/\Delta k=10\mu$m. Other parameters: $E_p=-4.75$meV, $V=5.1$meV,
$\Gamma=0.1$meV, $\zeta=0.025$, $\tau_\chi=50$ps, $\tau_\phi=3.3$ps,
$\alpha_1=5\times10^{-5}$meV mm$^2$, $\alpha_2=-0.1\alpha_1$,
$m_\chi=0.22m_e$ and $m_\phi=10^{-5}m_e$.}
    \label{fig:absphi}
    \end{figure}

    \begin{figure}[t]
        \centering
        \includegraphics[width=8.116cm]{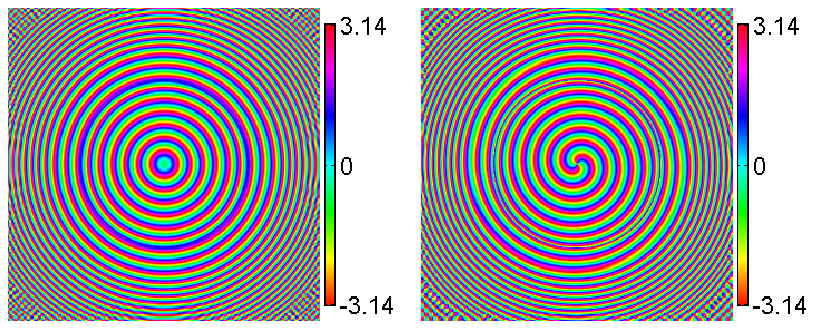}
    \caption{({\it colour}) Plots of the phase of the photonic
    wavefunction, $\arg(\phi_i)$, at 7.5ps after the arrival of the pulse for the co-circular (left) and cross-circular polarizations.}
    \label{fig:phasephi}
    \end{figure}

Equations (\ref{eq:GPchi}) and (\ref{eq:GPphi}) completely determine
the dynamics of interacting polaritons once initial wavefunctions
and pumping terms are defined. We acknowledge that the polariton
dynamics from a linear pulse pump was recently
studied~\cite{Shelykh2006} using similar equations (and our model
was able to reproduce these earlier results). In this paper we focus
on the case of excitation with a circularly polarized pump pulse
($A_+=1,A_-=0$) with broad spread in wavevector tuned slightly
($0.35$meV) above the bottom of the lower polariton branch. Although
the pump is centered at $\vec{k}_0=0$ (normal incidence) it is
resonant with the lower polariton branch along a ring
($\vec{k}=435$mm$^{-1}$) in reciprocal space with a radius set by
the pump energy. We numerically solved Eqs. (\ref{eq:GPchi}) and
(\ref{eq:GPphi}) to find the results plotted in Figs.
\ref{fig:absphi} and \ref{fig:phasephi}. We show the behaviour of
the photonic wavefunctions, which are directly accessible
experimentally. The excitonic wavefunctions demonstrate an identical
behaviour.

The appearance of expanding rings in the distributions should be
expected since we are exciting a ring in reciprocal space. In the
cross-circular polarization there is again a phase singularity and
vortex of winding number 2 (in agreement with the analytical linear
model).

    \begin{figure}[h]
        \centering
                \includegraphics[width=8.116cm]{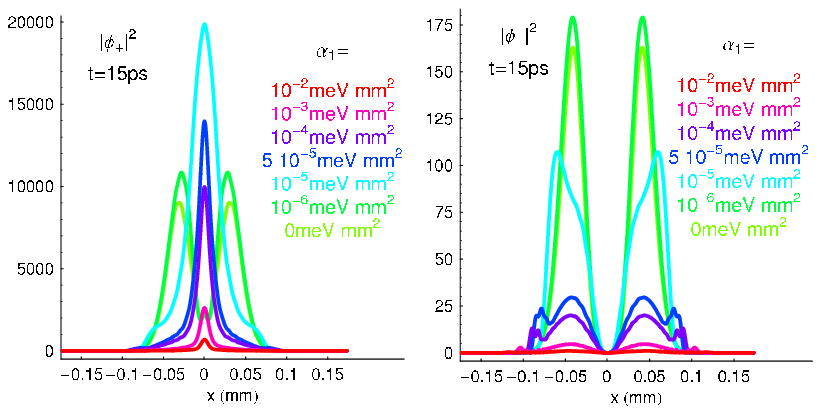}
    \caption{({\it colour}) The intensity of the photonic wavefunction,
    $|\phi_i|^2$, in the co-circular (left) and cross-circular
    (right) polarizations for different values of $\alpha_1$ at $15$ps after the peak of the excitation
    pulse.  We keep the relationship $\alpha_2=-0.1\alpha_1$.}
    \label{fig:U0dependence}
    \end{figure}

The expanding polariton rings are influenced significantly by the
polariton-polariton interactions as shown in Fig.
\ref{fig:U0dependence} where we plot cross sections of the
wavefunctions along the x-axis for different values of the
scattering parameter, $\alpha_1$. In the absence of
polariton-polariton interactions ($\alpha_1=0$) a ring is observed
in both circular polarizations. As $\alpha_1$ increases the
intensity of the wavefunctions increases slightly (formally an
increase in the interaction constants is identical to an increase in
the pump amplitude) but as the interaction constants increase
further the intensity decreases, and the wavefunctions spread due to
the repulsion of polaritons in real space. Also note that as
$\alpha_1$ increases we observe a central peak in the co-circular
distribution instead of a ring. This is due to the larger
(time-dependent) blueshift of the system when $\alpha_1$ is
increased, which results in the resonant excitation of a smaller
ring in reciprocal space. In real space we also observe the
appearance of darker rings within the brighter ring in
cross-circular polarization as $\alpha_1$ is increased from zero.
Such structures seem similar to dark ring solitons, which are known
to appear in the equilibrium scalar Gross-Pitaevskii equations and
similar non-linear Schr\"odinger
equations~\cite{Theocharis2003,Kivshar1994RDS}. The incoherent
polariton scattering, introduced via dissipative terms in Eqs.
(\ref{eq:GPphi}) and (\ref{eq:GPchi}), does not affect the
appearance of the vortices, whilst it does influence the intensity
of polarized light emitted by the cavity.


{\bf Conclusion.} We calculated the dynamics of an interacting
polariton condensate, which was excited by a circularly polarized
Gaussian pump pulse tuned above the bottom of the lower polariton
branch. The presence of TE-TM splitting in the microcavity allows
coupling into the cross-circular polarization in which the phase
shows the profile of a vortex of winding number 2. The appearance of
a linear polarization vortex and spreading ring-like structures is
also derived.


The authors thank M. M. Glazov, Yu. G. Rubo, G. Malpuech and J.
Ruostekoski for helpful comments. T.C.H.L acknowledges support from
the E.P.S.R.C.


\end{document}